# In search of Majorana

Sankar Das Sarma[1]

[1] Condensed Matter Theory Center and Joint Quantum Institute,

Department of Physics, University of Maryland, College Park, MD 20742, USA

**Majorana particles are the same as their antiparticle, and their analogues in condensed matter may be a platform for quantum computing. We describe the search for these modes in semiconductor heterostructures and how disorder is a limiting factor.**

Ettore Majorana was a brilliant Italian theoretical physicist, who suddenly vanished one day in 1938 leaving no trace, right after finishing his most famous work associated with Majorana fermions. [1] Rumor had it that he fell into the sea and drowned or jumped into the water committing suicide or retired as a reclusive monk in a monastery or perhaps moved to South America opening a café in secrecy somewhere. No one knows what happened. This Perspective discusses Majorana's profound effect on condensed matter physics and in particular the attempts to observe a version of the particles he predicted in hybrid structures of superconductors and semiconductors.

Majorana postulated that in principle there could be real solutions to the Dirac equation associated with a new type of neutral fermion that is its own anti-particle. Such a creation operator, the Majorana operator, is then self-adjoint by definition since the particle and the anti-particle are the same. Neutrinos may or may not be Majorana fermions – ongoing experiments all over the world are trying to figure out if they are, with huge implications for fundamental physics.

In condensed matter physics, we cannot create new particles on demand, we only have electrons. But self-adjoint Majorana operators can arise naturally in two-dimensional electron systems, creating emergent localized excitations where the particles are their own antiparticles. This is appealing, because these Majorana excitations are not fermions, but are topological objects called 'non-Abelian anyons'. These excitations have exchange statistics that are neither Fermi nor Bose statistics. [2] This is allowed in two dimensions because exchanging particles involves the braid group, rather than the permutation group as in three dimensions.  The most famous examples of such anyons occur in fractional quantum Hall effects where the excitations could be either abelian (for example the 1/3 state) or non-abelian (5/2 state).  It is predicted that these emergent Majorana excitations can be braided around each other in well-defined manner to carry out fault-tolerant topological quantum computation, providing a huge practical incentive to search for them. As an aside, neutrinos, even if they turn out to be Majorana fermions, are rather pedestrian in that they are neither topological nor anyonic, they are just neutral Dirac fermions whose creation operators are self-adjoint.  One cannot do topological quantum computing with neutrinos.

It was realized some time [3,4] ago that superconductors are natural hosts for Majorana particles because the in-gap excitations of superconductors, called Bogoliubov excitations, are created by composite operators consisting of a linear combination of electrons and holes arising from an exact electron-hole symmetry inherent to superconductors.[3,4] All that is needed is to find the precise neutral excitation that is an equal linear combination of electron and hole operators. By construction, this operator is a self-adjoint Majorana operator that creates mid-gap neutral excitations that are neither pure electron nor pure hole. In fact, such 'zero energy' (meaning they sit right at the center of the superconducting gap) excitations are effectively half an electron or half a hole, which must be anyonic. Given that in condensed-matter physics we only have either electrons or holes, an immediate consequence is that such Majorana zero-mode (MZM) excitations can never emerge in isolation by themselves, they must always come in pairs so that two of them together is an electron or a hole or nothing.

Superconductors hosting midgap MZMs are called 'topological superconductors', and the simplest example of such a Majorana-hosting topological superconductor is a spinless orbital *p*-wave superconductor.[4] The issue for MZMs and topological quantum computing appears to be straightforward:  Just find a two-dimensional spinless *p*-wave superconductor, and look for the midgap non-Abelian localized Majorana zero modes, braid them, and create a fault-tolerant topological quantum computer. After all there are thousands of superconductors of all possible orbital symmetries so surely some of them must be spinless or can be rendered spinless easily.

Alas, it does not seem that spinless *p*-wave superconductors exist at all in nature, never mind two-dimensional ones. So, an alternative is to try to engineer them in heterostructures. Between 2006 and 2009, theorists worked hard and came up with several suggestions [5-11], some reasonable and some less so, to create laboratory systems that work like two-dimensional spinless *p*-wave superconductors under certain conditions. In particular, Ref. 8 introduced the physical idea of proximity superconductivity as a tool in this context, and Refs. [8-11] emphasized that spin-orbit coupling may help.   Some of these suggestions, particularly the ones involving fractional quantum Hall states, are still worth pursuing.

Then, in 2009, the Maryland group wrote a paper that proposed an experimentally-tractable alternative for an artificially engineered effective spinless *p*-wave superconductor in semiconductor/superconductor (SM/SC) hybrid structures. [12] The precise proposal that got the experimentalists' attention is a one-dimensional (1D) adaptation [13-16] of previous ideas in two-dimensional systems with an applied magnetic field that could be used to spin-polarize the system [16].

MZMs are, by definition, localized modes at the boundary of the system, so one qualitative difference between one and two dimensions is that MZMs in 1D are completely localized quantum bound sates (almost like an effective 'particle') at the ends of the 1D chain. This 1D semiconductor-based realistic Majorana proposal is in some sense an experimentally-realizable version of an older idealized 1D tight-binding lattice spinless *p*-wave superconductor model [17]. By construction such a model contains end MZMs, and could carry out topological quantum computing by suitably braiding them. [17-19]

The basic idea [13-15] is simple (Figs.1 and 4): Take a 1D semiconductor nanowire (made from, for example, InAs or InSb) that has sufficiently strong spin-orbit coupling and put it in close contact with a regular metallic superconductor (for example aluminium or niobium) so that the wire develops proximity induced superconductivity from the parent superconductor. [8] Then, apply a magnetic field parallel to the wire length to create a Zeeman splitting that lifts the degeneracy of opposite spins. When the Zeeman splitting is strong enough (see Fig. 4), one of the spin types becomes effectively inactive, lifting the spin degeneracy, and the nanowire will host a spinless $p$-wave topological superconducting state (Fig. 2).

One significant advantage of this engineered 1D Majorana construction using SM/SC hybrid structures is its tunability by varying either the applied magnetic field (to change the Zeeman splitting) or the voltages applied on various gates (to change the chemical potential). This can tune the device in and out of the topological regime. If the wire is long (much longer than the coherence length), the two end MZMs that exist in the topological phase are effectively isolated, and act as non-Abelian anyons that are immune to all local perturbations that preserve fermion parity and are protected by the topological gap itself.

Although the existence of the critical Zeeman splitting where the nanowire superconducting gap opens or closes is a necessary condition for the topological quantum phase transition (TQPT) and the associated MZMs, no published experiment has as yet definitively reported this feature. The experimental focus has been on the predicted midgap tunneling transport through the MZMs, which should produce a zero bias conductance peak (ZBCP), meaning a finite differential conductance at zero voltage, in local tunneling spectroscopy data from both ends of the wire. [20-23]

In 2012, interest and excitement in the subject grew enormously as several experimental groups reported [24-28] the observations of ZBCPs in InSb or InAs nanowires in the presence of Al or Nb as the parent superconductor under an applied magnetic field of about one Tesla. In solid state physics involving complex materials, it is very rare for five different well-known experimental groups to apparently verify the same theoretical predictions more or less simultaneously, as happened here.  There was euphoria over the physics, and the strange mystery of the disappearance of Ettore Majorana served as a romantic background in spite of these localized MZMs really having little to do with Majorana's old work.

There was also an additional experimental report in 2012 of the observation of the theoretically predicted [29, 13] fractional Josephson effect in InSb/Nb system [30]. Theoretically this arises from the qualitative fact that the MZM is effectively half an electron leading to an effective  *h/e* 'fractional' Josephson effect, compared with the standard *h/2e* effect, thus doubling the flux periodicity.  This experiment has never been reproduced in contrast to the ZBCP observations in tunneling, which have been reproduced many times by many groups.  With so many groups reporting MZMs in SM/SC structures under an applied magnetic field, the community started to believe that MZMs had been seen, and a MZM-based topological quantum computing would soon follow.  There were many exaggerated popular articles and many immodest press releases declaring Majorana victory far too soon.

Alas, these early claims (and subsequent published experimental claims of Majorana sighting in nanowires) turned out to be hasty and incorrect. However, the silver lining to that cloud was that these experiments represented considerable scientific progress, advancing our understanding of the subtle and complex physics of SM/SC hybrid systems.  One significant accomplishment of MZM research during 2012-2016 is the achievement of the hard proximity SC gap in SM nanowire by using epitaxial SM/SC systems [31] following specific theoretical predictions [32], in contrast to the soft proximity gap [24-28] with substantial subgap fermionic states in the early (2012-2013) experiments.

With the benefit of hindsight, among the many problems that should have been obvious in the 2012-2013 experiments are that the wires they were performed on were not in the long-wire limit, the induced SC gap was very soft with obvious subgap states, and there was a complete absence of any gap closing and reopening that is essential for a TQPT. There was also an absence of any end-to-end nonlocal correlations between tunneling carried out from the two ends as there must be for the non-local MZMs, an absence of any stability against parameter changes, an absence of any Majorana oscillations with increasing field because of the overlap of the MZM wave functions, and very poor statistics of the occurrence of ZBCPs.

So, why were the Majoranas not there? We know now [33-39] that disorder was playing a dominant role, and what was observed are signatures of disorder-induced non-topological subgap fermionic Andreev bound states (ABSs) rather than anyonic topological MZMs. A disorder-induced ABS could once in a while be rather close to zero energy, thus accidentally simulating some features (for example, the ZBCPs) of MZMs.

My opinion [37-39] is that ABSs were mimicking some, but not all, of MZM properties occasionally and superficially, and with sufficient fine-tuning and post-selection of the data the experimentalists could always find a small fraction of their data in a small fraction of their samples that looked misleadingly like 'MZM signatures'; most particularly ZBCPs almost at zero-energy may arise naturally from disorder-induced ABSs in a finite magnetic field. Essentially, the claims of MZM signatures (or MZM observations) were, in my view, all results of honest and unwitting confirmation bias, arising from the fact that the theoretical predictions are so precise that it was easy to be misled that they have been experimentally verified. An unfortunate exclusive focus on ZBCP observations in the local tunneling spectroscopy (rather than, for example, nonlocal signals or gap reopening or Majorana oscillatory features) had led the experimentalists down the wrong track.

The fact that disorder may be playing a role was already pointed out [40-43] in some early cautionary theoretical papers during 2012-13, but their key importance became manifestly clear with detailed analyses [35, 36] of two important experimental papers in 2016. One, by Albrecht et al [44] claimed the observation of the predicted topological protection arising from the MZM separation being larger than the coherence length so that the MZM wave function overlap from the two ends is exponentially small. The other by Deng et al [45] claimed to observe the merging of two ABSs into the MZM leading to the ZBCP. Although no theory can definitively prove that MZMs were not seen in these (or any other) measurements, it nonetheless established compellingly that in the presence of disorder, experimental signatures involving ZBCPs are a necessary but not sufficient condition to claim the existence of MZMs.

More recently, the Maryland group carried out a detailed theoretical analysis [37] of all SM/SC tunneling experiments arguing that there can be three different types of ZBCPs: good (actual MZMs), bad (ABSs, sometimes called quasi-MZMs, produced by accidental quantum dots in the nanowire), ugly (produced by random disorder induced ZBCPs). The conclusion is that, most likely, all SM/SC structures have so far observed only disorder-induced ugly ZBCPs although occasionally bad ZBCPs may also have manifested, but good ZBCPs arising from MZMs have not been seen.

In particular, the good/bad/ugly analysis [37] specifically addressed the experimental claims of the observation of 'Majorana quantization', meaning that the ZBCPs have a precise value of $2e^2/h$ as appropriate for perfect MZMs at $T$=0. [21-23] Occasionally disorder-induced ZBCPs could manifest approximate $2e^2/h$ apparent 'quantization', and experiments most likely reported these ugly ZBCPs through fine-tuned post-selection (Fig. 3). Our 'ugly' simulations [37] generically seem to produce fine-tuned and post-selected ZBCPs indistinguishable from the experimental data.

This work predated the unpleasant controversy around the reproducibility of MZMs [46] that arose leading to the retraction of the Zhang Nature 2018 article (see H. Zhang et al., Nature **556**, 74 (2018)). In fact, the updated version of that paper following the retraction [47] concedes that the observed approximate $2e^2/h$ ZBCPs are most likely disorder-induced ugly peaks discussed in [37], settling the technical aspects of this controversy. The decisive role of confirmation bias in the retracted Nature paper has been emphasized by a report from a group of independent experts appended to the retraction note. The good/bad/ugly paper [37] also showed that the claim of the MZM observation [48] based on the temperature scaling is completely consistent with disorder-induced ugly ZBCPs.

Our realistic numerical modeling and theoretical analysis [49, 50] of the experimental samples convince me that the current SM/SC samples are roughly 10-100 times dirtier than what would be necessary to realize pristine 'good' ZBCPs arising from topological MZMs. But what is this disorder and where is it located?

The main disorder is in the SM nanowire itself and at the SM/SC interface, and arises from the parent SM material being not clean enough.  We also know from numerical simulations, compared with both transport and tunneling data that the disorder essentially arises from unintentional random charged impurities in the semiconductor. [49, 50] This materials problem should be solvable by improving semiconductor growth techniques—it is not cheap to do so, but developing a new technology and observing completely new types of 'particles' cannot be either easy or inexpensive.

The mechanism by which the disorder compromises the topological superconductivity is also known. Remember that two completely overlapping MZMs are basically fermions, and two nearby MZMs are equivalent to ABS. Only MZMs very far from each other with little overlap are non-Abelian anyons.  While topological SC is immune to weak disorder, very strong disorder necessarily mixes MZMs, generically producing ABSs that can occasionally mimic some MZM properties.  Also, in short wires there are no MZMs since their overlap is strong. Current published experiments are all in short-wire and strong-disorder regimes.

I fully expect topological MZMs to be observed in the near future in SM/SC nanowires, once the samples are made purer.  The underlying theory is a free fermion band structure theory which should essentially be exact for the experimentally studied SM/SC systems, except for the problems of disorder and short wires.  Just as our current Si-based CMOS electronics industry depends crucially on highly purified Si wafers through extremely complex engineering, the eventual Majorana technology will depend on ultra-pure InAs nanowires on Al substrates. Once the nanowires are clean enough, true MZMs will emerge in long wires, and the condensed matter version of the search for Majorana will end.

So, materials improvement would lead to MZMs, and this appears to be the current emphasis of the community working on MZMs in SM/SC structures. What matters is the dimensionless ratio of the disorder strength to the pristine topological SC gap, which cannot be much larger than unity for MZMs to emerge. Currently, this ratio is of the order of 10 or 100. Therefore, future progress must involve reducing disorder in the currently used InAs/Al samples, or finding new materials where the induced SC gap in the SM is much larger, so that the need for reducing disorder is less stringent. The future is bright because we know what the problems are and how to solve them. The problems are difficult, but by no means insurmountable.

I will be convinced once experiments in ultra-pure SM/SC systems observe simultaneous stable correlated ZBCPs generically from both ends in local tunneling (without extreme fine-tuning) in nanowires that are longer than the coherence length along with the controlled closing and opening of a gap in nonlocal tunneling, indicating the TQPT. A very recent experiment from Microsoft might indicate such results, but the situation is unclear since the sample still has considerable disorder, and the claimed topological gap is tiny. [51] More work is obviously necessary to decisively resolve the situation. Once MZMs are seen, braiding experiments would show that 'our' condensed matter Majorana is much more interesting than anything even the genius of Ettore Majorana could have ever imagined.

**Acknowledgment**

The author thanks Dr. Haining Pan for help with the manuscript preparation. The author also thanks Drs. Maissam Barkeshli, Haining Pan, Jay Deep Sau, and Tudor Stanescu for reading and commenting on the manuscript. The author's work is supported by the Laboratory for Physical Sciences.

**REFERENCES**

1. E. Majorana, Il Nuovo Cimento **14**, 171–184 (1937)
2. C. Nayak, *et al.*, Rev. Mod. Phys. **80**, 1083-1159 (2008)
3. C. Beenakker, Phys. Rev. Lett. **112**, 070604 (2014)
4. N. Read and D. Green, Phys. Rev. B **61**, 10267-10297 (2000)
5. S. Das Sarma, M. Freedman, and C. Nayak, Phys. Rev. Lett **94**, 166802 (2005)
6. S. Das Sarma, C. Nayak, and S. Tewari, Phys. Rev. B 73, 220502 (2006)
7. S. Tewari*, et al*., Phys. Rev. Lett. **98**, 010506 (2007)
8. L. Fu and C. Kane, Phys. Rev. Lett. **100**, 096407 (2008)
9. C. Zhang, *et al*., Phys. Rev. Lett. **101**, 160401 (2008)
10. M. Sato, Y. Takahashi, and S. Fujimoto, Phys. Rev. Lett. **103**, 020401 (2009)
11. M. Sato and S. Fujimoto, Phys. Rev. B **79**, 094504 (2009)
12. J. Sau, *et al.*, Phys. Rev. Lett. **104**, 040502 (2010)
13. R. Lutchyn, J. Sau, and S. Das Sarma, Phys. Rev. Lett. **105**, 077001 (2010)
14. Y. Oreg, G. Refael, and F. von Oppen, Phys. Rev. Lett. **105**, 177002 (2010)
15. J. Sau, *et al*., Phys. Rev. B **82**, 214509 (2010)
16. J. Alicea, Phys. Rev. B **81**, 125318 (2010)
17. A. Kitaev, Physics Uspekhi **44**, 131–136 (2001)
18. A. Kitaev, Annals Phys. **303**, 2-30 (2003)
19. S. Bravyi and A. Kitaev, Annals of Physics **298**, 210-226 (2002)
20. K. Sengupta, *et al.*, Phys. Rev. B **63**, 144531 (2001)
21. K. Flensberg, Phys. Rev. B **82**, 180516 (2010)
22. K. T. Law, P. A. Lee, and T. K. Ng, Phys. Rev. Lett. **103**, 237001 (2009)
23. M. Wimmer, *et al.*, New J. Phys. **13**, 053016 (2011)
24. M. Deng, *et al*., Nano Letters **12**, 6414-6419 (2012)
25. V. Mourik, *et al*., Science **336**, 1003-1007 (2012)
26. A. Das, *et al.*, Nature Physics **8**, 887 (2012)
27. H. Churchill, *et al*., Phys. Rev. B **87**, 241401 (2013)
28. A. Finck, *et al*., Phys. Rev. Lett. **110**, 126406 (2013)

29. H. Kwon, K. Sengupta, and V. Yakovenko, Eur. Phys. Journal B 37, 349-361 (2004)
30. L. Rokhinson, X. Liu, and J. Furdyna, Nat. Phys. **8**, 795-799 (2012)
31. W. Chang, *et al*., Nat. Nano. **10**, 232-236 (2015)
32. S. Takei, *et al.,* Phys. Rev. Lett. **110**, 186803 (2013)
33. J. Sau, S. Tewari, and S. Das Sarma, Phys. Rev. B **85**, 064512 (2012)
34. J. Sau and S. Das Sarma, Phys. Rev. B **88**, 064506 (2013)
35. C. Liu, *et al*., Phys. Rev. B **96**, 075161 (2017)
36. C. Chiu, J. Sau, and S. Das Sarma, Phys. Rev. B **96**, 054504 (2017)
37. H. Pan and S. Das Sarma, Phys. Rev. Res. **2**, 013377 (2020)
38. S. Das Sarma and H. Pan, Phys, Rev. B **103**, 195158 (2021)
39. H. Pan, *et al*., Phys. Rev. B **101**, 024506 (2020)
40. G. Kells, D. Meidan, and P. Brouwer, Phys. Rev. B **86**, 100503 (2012)
41. A. Akhmerov, *et al*., Phys. Rev. Lett. **106**, 057001 (2011)
42. J. Liu, *et al*., Phys. Rev. Lett. **109**, 267002 (2012)
43. D. Bagrets and A. Altland, Phys. Rev. Lett. **109**, 227005 (2012)
44. S. Albrecht, *et al.,* Nature **531**, 206 (2016)
45. M. Deng, *et al.,* Science **354**, 1557–1562 (2016)
46. S. Frolov, Nature **592**, 350-352 (2021)
47. H. Zhang, *et al*., arXiv:2101.11456 (2021)
48. F. Nichele, *et al*., Phys. Rev. Lett. **119**, 136803 (2017)
49. S. Ahn, *et al*., Phys. Rev. Materials 5, 124602 (2021)
50. B. D. Woods, S. Das Sarma, T. D. Stanescu, Phys. Rev. Applied 16 054053 (2021)
51. M. Aghaee *et al*., arXiv:2207.02472 (2022)
52. C. Quay *et al*. Nature Physics **6**, 336-339 (2010)

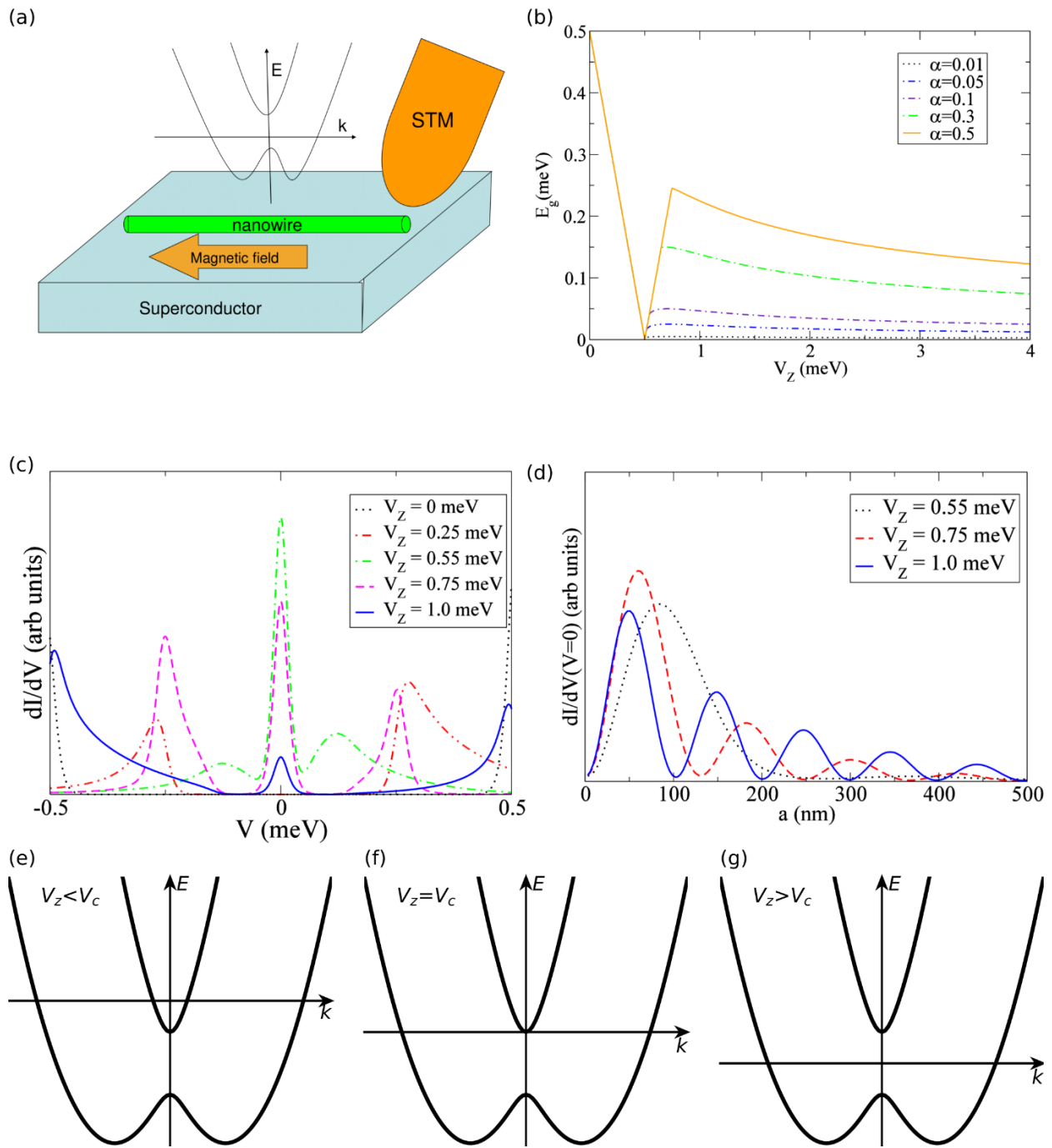

**FIGURE 1 | Theoretical predictions for experimental signatures of MZMs.** (a) Proposed SM/SC experiment with nanowire, superconductor, magnetic field, tunneling contacts (inset: schematic band structure showing effects of SO coupling and Zeeman splitting); (b) Shows the calculated induced gap closing and reopening as a function of the Zeeman splitting $V_z$. The topological quantum phase transition occurs at the critical value when $V_z = V_c = 0.5$ meV; (c) Calculated

tunnel conductance spectra as a function of bias voltage for different values of $V_z$ (with $V_c$ = 0.5 meV) showing MZM-induced ZBCPs for $V_z > V_c$; (d) Calculated ZBCP strength for $V_z > V_c$ (= 0.5 meV) as a function of wire length showing Majorana oscillation becoming more prominent for shorter wires. The bottom panels show the schematic bands for three situations: (e) trivial, (f) TQPT (assuming the proximitized superconducting paring energy is small), (g) topological. (Panels a-d: adapted from Ref. [15]; panels e-f: Courtesy of Haining Pan)

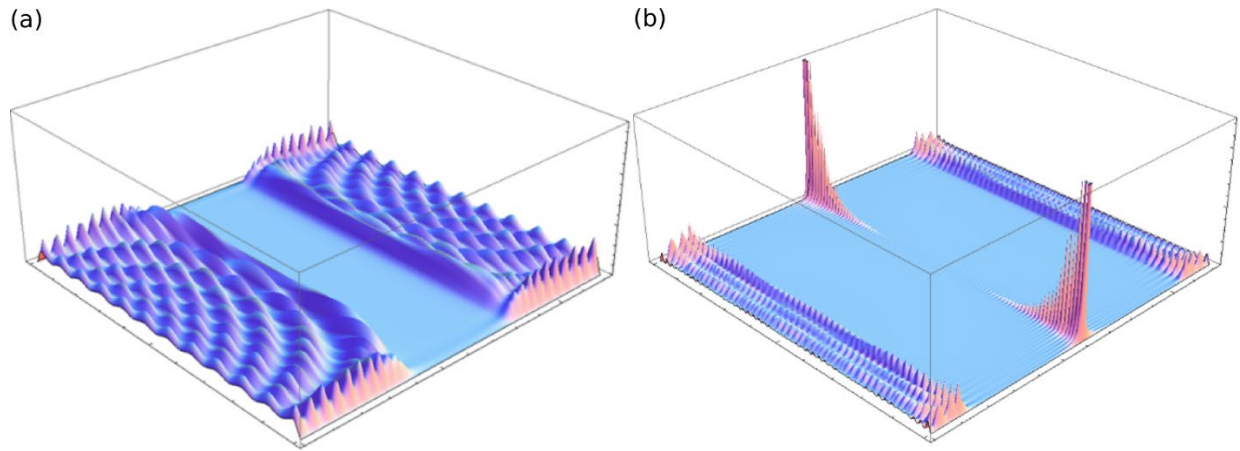

**FIGURE 2 | Excitation spectra in the topological and trivial regimes.** (a) Numerically calculated excitation spectra in the nanowire for (a) $V_z < V_c$ (where $V_c$ is the Zeeman splitting that corresponds to the TQPT) with no states in the superconducting), and (b) for $V_z > V_c$ showing localized MZMs at the wire ends. There are no states in the gap other than the two MZMs localized at the wire ends for $V_z > V_c$, whereas there are continuum metallic electron-hole excitations above the SC gap. (Courtesy of Tudor Stanescu)

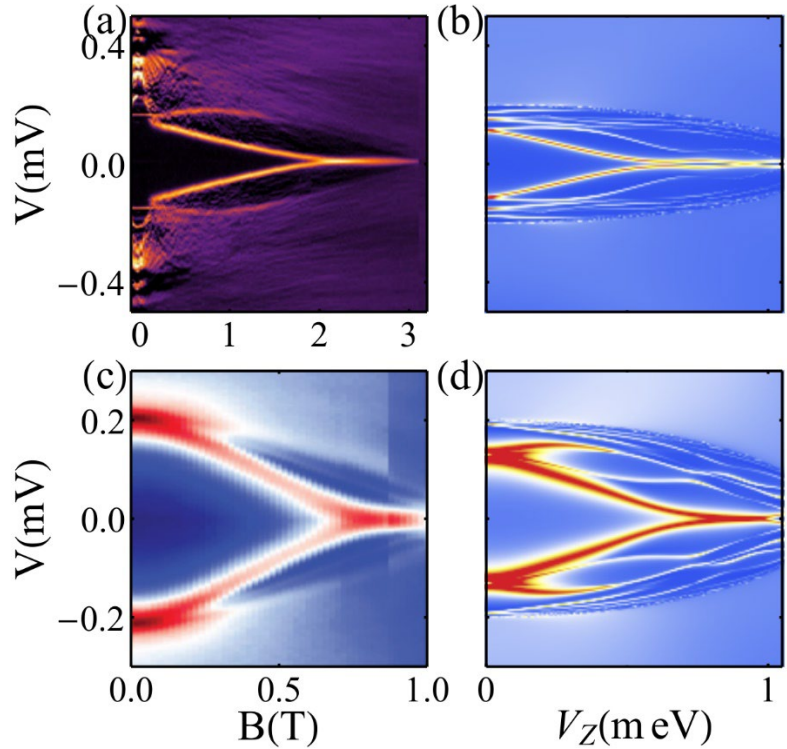

**FIGURE 3 | The role of disorder in nanowire experiments**. The calculated trivial or 'ugly' tunnel conductance spectra in realistic disordered SM/SC nanowires (right column) compared directly with the corresponding experimental results (left column). The data in the top row comes from Ref [48], and that in the lower row from Ref [47]. (Figure adapted from Ref. [37].)

**Appendix**

The key idea is that if spinless *p*-wave superconductors do not exist in nature, we must engineer one in the laboratory by combining well-known and easily available components that, when appropriately combined in a hybrid structure, will manifest this state under well-defined conditions of parameter tuning. The necessary components are superconductivity and spinlessness. Gaining a spinless superconductor sounds easy – just apply a magnetic field that polarizes it. But this does not work because spin splitting would simply suppress the superconducting gap to zero when the spin splitting equals the gap. This is the well-known Pauli limit or Clogston effect, where spin polarization suppresses the gap in an *s*-wave superconductor, making all superconductivity vanish. Instead, what is necessary is to create a spin texture at the Fermi surface along with a spin splitting so that, with increasing spin splitting, the system can retain its superconductivity. These can be done via spin-orbit coupling (Fig. 4).

Since a spinless superconductor must have the spin part of its wave function being symmetrical, its orbital part must be antisymmetric to ensure that the overall wave function is antisymmetric. This implies the orbital wave function must be *p*-wave (meaning it has angular momentum of 1) or *f*-wave (angular momentum of 3). We know how to create this spin texture using spin-orbit coupling that is native to many semiconducting materials, and inducing superconductivity via the proximity effect.

The theory provides a crisp prediction for the development of an effectively spinless *p*-wave topological superconductivity in the nanowire:

$$V_z > V_c = \sqrt{\Delta^2 + \mu^2},$$

where $V_z$ is the field-induced Zeeman spin splitting in the nanowire, $\Delta$ is the zero-field proximity-induced superconducting gap in the nanowire, and $\mu$ is the chemical potential in the semiconductor. The field $V_c$ defines the topological quantum phase transition (TQPT) separating trivial from topological.

When $V_z < V_c$, the system is in the trivial (meaning non-topological) phase and for $V_z > V_c$, it is in the topological phase with Majorana zero modes at the wire ends. (Fig.2) At the 'critical field' $V_c$, the induced effective gap in the nanowire vanishes (this is the same as the Pauli limit described above), characterizing a TQPT from an ordinary spinful (*s*-wave) trivial superconductor ($V_z < V_c$) to a topological effectively spinless *p*-wave superconductor ($V_z > V_c$), with a topological gap opening for $V_z > V_c$ accompanied by the appearance of midgap MZMs as defect states localized at the two wire ends. (Fig. 1) The topological gap is proportional to the spin-orbit coupling strength, which does not affect $V_c$ itself.

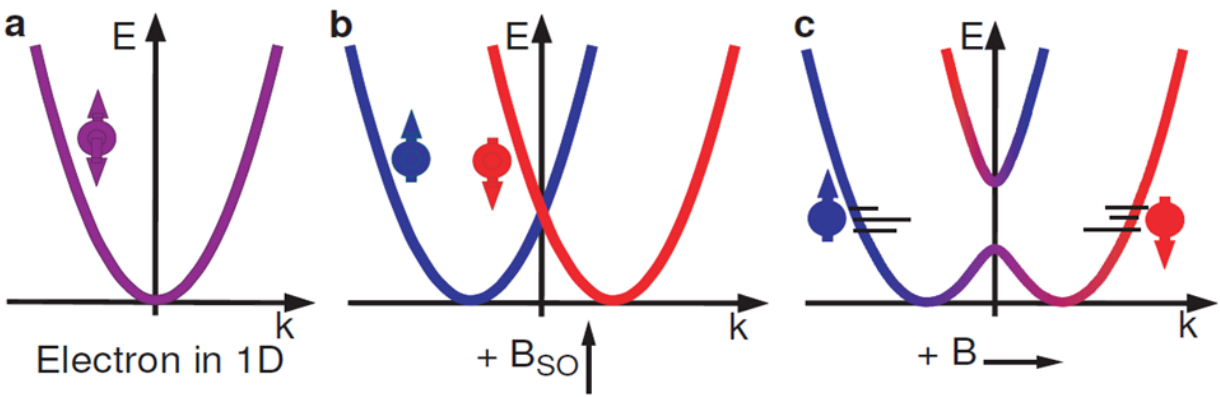

**FIGURE 4** The basic idea for engineering a 1D spinless p-wave SC in SM/SC nanowires, using spin-orbit coupling $B_{so}$ (which shifts the spin- up/down levels in momentum space) and Zeeman effect *B* (which creates a spin splitting), thus leading to a spin texture at the Fermi level. A proximity-induced SC gap inside the spin split levels would lead to an effective spinless *p*-wave SC induced by *s*-wave SC + spin-orbit coupling + Zeeman spin splitting. (Figure adapted from Ref. [52].)